\documentclass[epj]{svjour1}

\usepackage{epsfig}

\def \as {\relax\ifmmode\alpha_s\else{$\alpha_s${ }}\fi}
\def \LQCD {\Lambda_{\mbox{\tiny QCD}}}

\def\ba {\begin{eqnarray}}
\def\ea {\end{eqnarray}}

\def\be {\begin{equation}}
\def\ee {\end{equation}}

\begin{document}

\title{Power Corrections to $\mathbf{e^+ e^-}$ Dijet Event Shapes\,\thanks{based on talk given by C.F.\ Berger at EPS HEP 2003, Theme: ``Hard QCD'', Aachen, Germany, July 2003.}}

\author{Carola F. Berger\inst{1} 
\and George Sterman\inst{2}
}                     
\institute{I.N.F.N. Sezione di Torino, Via P. Giuria 1, I-10125 Torino, Italy
%, \email{carola.berger@to.infn.it}
\and 
C.N.\ Yang Institute for Theoretical Physics, Stony Brook University, Stony Brook, NY 11794-3840, U.S.A.
%, \email{sterman@\-insti.\-physics.sunysb.edu}
}

%\date{Received: date / Revised version: date}
\date{\today}

\abstract{We discuss a class of event shapes for $e^+e^-$ dijet events that include the thrust as a special case. Large logarithmic corrections to the corresponding cross sections can be resummed to all logarithmic orders at leading power. However, irrespective of the order up to which the perturbative expansion is calculated, it has to be supplemented by nonperturbative corrections due to its at best asymptotic nature. We find that the leading power corrections are universal for the class of event shapes discussed here. Based on these findings, we provide sample numerical predictions for the distributions of the new event shapes. 
} 
\maketitle

\section{Motivation}
\label{intro}

Event shapes are generalizations of jet cross sections that describe the distribution of radiation in the final state \cite{shapes}. They are thus sensitive to both the underlying short-distance hard scattering and to long-time hadronization processes. In the following, although not always stated explicitly, we will consider event shapes in $e^+ e^-$ annihilation, where long-time effects only affect the final state.

The hard scattering is responsible for the basic form of the distribution of radiation. 
For example, the contribution of 
a two-jet event to an event shape cross section  is distinctly different from that of a three-jet 
final state. 
The distribution of radiation stemming from the short-distance process 
can be computed within perturbation theory in terms of quarks and gluons. Originally, this was one of the main motivations to study event shapes, to test QCD \cite{shapes}. 

The correctness of QCD was quickly established by confirming the main features of event shape distributions \cite{shapes}. Despite this success, perturbation theory cannot fully account for their exact form. This is due to hadronization effects which enter as nonperturbative (NP) power corrections. The main, although not only, effect of the hadronization process is to widen the distribution of radiation in the final state, and thus to shift the peak of the event shape distribution away from the narrow-jet limit \cite{irrdiff,DokWeb}. For the description of this shift it suffices to consider the first power correction, typically proportional to $1/Q$, where $Q$ is the center of mass (c.m.) energy. In general, this gives an accurate picture of the energy dependence of the average values of a variety of $e^+ e^-$ event shapes with only a small set of parameters that are fit to experiment (see \cite{alphas} and references therein). For differential distributions, however, one needs to take into account more than the first power correction, which leads to the introduction of nonperturbative shape functions \cite{shapefn}. Perturbation theory imposes constraints on these shape functions such that good agreement with experiment can be achieved, again with only a small set of NP parameters \cite{applshape}.

The present state of the art is perturbative computations up to next-to-leading order (NLO) in the strong coupling, with large logarithmic corrections due to soft and/or collinear radiation resummed to next-to-leading logarithm (NLL). Supplemented with the aforementioned set of nonperturbative parameters, many $e^+ e^-$ event shape cross sections can be predicted with impressive accuracy. 
One of the main applications is thus the precise determination of the strong coupling \cite{alphas}. Nevertheless, as we will illustrate below, there is still more to be learned from the theoretical study of such event shapes, aside from the computation of yet higher orders or their numerical evaluation.

In Ref. \cite{BKuS}, the authors, together with T. K\'ucs, 
introduced a class of event shapes $\tau_a$, depending on a continuous, real, parameter $a$, 
that includes as special cases the familiar
thrust \cite{Farhi:1977sg} and jet broadening \cite{broad1}. As we will review below in Section \ref{sec:resum}, the introduction of this parameter $a$ allows us to analytically study a whole range of event shapes simultaneously, and from its variation deduce a variety of consequences for both the perturbatively calculable contributions \cite{BKuS}, and the long-distance corrections \cite{BS1,BS}. The factorization of long- and short-distance effects results in a scaling rule that relates the power corrections within the considered class of event shapes, which is described in  Sec. \ref{sec:nonpt}. We provide numerical examples, and conclude by summarizing present and possible future perspectives.  

For sake of brevity, we only quote here the main features of the analytical results obtained in Refs. \cite{BKuS} and \cite{BS}. More details, an account of various technicalities, and further references can be found there. Here we show some new numerical results.

\section{Resummation of Large Logarithms}
\label{sec:resum}

We consider the following distributions of 
event shapes $\tau_a$ that weight final states $N$ in $e^+ e^-$ annihilation processes in the narrow two-jet limit, 
\be
\frac{d \sigma(\tau_a, Q)}{
d\tau_a}
=
{1\over 2 Q^2}\ \sum_N\;
|M(N)|^2\,  \delta(\tau_a - \tau_a(N)),
\label{eventdef}
\ee
with a center of mass energy $Q \gg \LQCD$. We sum over all final states $N$ that contribute to the
weighted event. $M(N)$ denotes
the corresponding amplitude. 
The weight functions defined
in Ref.\ \cite{BKuS} for a state $N$ are 
\be
\tau_a(N)
= {1\over Q}\sum_{{\rm all}\ i\in N}\ p_{i\perp}\; e^{-|\eta_i|(1-a)}\, ,
\label{barfdef}
\ee
where $p_{i\perp}$ is the transverse momentum of particle $i$ relative to the 
thrust axis, 
and $\eta_i$ is the corresponding pseudorapidity, 
$\eta_i = \ln \cot \left(\theta_i/2\right)$. The thrust axis 
\cite{Farhi:1977sg} is
the axis with respect to which the above expression is minimized at $a = 0$.
Parameter $a$ is
adjustable, $- \infty < a < 2$, and allows us to study
various event
shapes within the same formalism.  
The case $a=0$ in Eq.\ (\ref{barfdef}) is essentially $1-T$, with $T$
the thrust while $a=1$ is the jet  broadening. 

In the two-jet limit, $\tau_a \rightarrow 0$, and  (\ref{eventdef}) has large corrections in $\ln \left(1/\tau_a\right)$ that have been resummed to all logarithmic orders at leading power for $a < 1$ in \cite{BKuS}. For $a \sim 1$, recoil effects have to be taken into account, as was pointed out for the broadening ($a = 1$) in Ref. \cite{Dokshitzer:1998kz}. 

In the following we will quote the result of the resummation of large
logarithms of $\tau_a$ in Laplace moment space: 
\ba
      \tilde{\sigma} \left(\nu,Q,a \right) & = &  \int^1_0 d \tau_a\, {\rm e}^{\;
-\nu\,
\tau_a}\ {d
\sigma(\tau_a,Q) \over d \tau_a}\, .
\label{trafo}
\ea
Logarithms of $1/\tau_a$ are transformed to logarithms of $\nu$. 

We refrain from quoting the slightly more complex formula for the resummed cross section valid to all logarithmic orders \cite{BKuS}, and only give the result at NLL,
\ba
 {1\over \sigma_{\rm tot}} \, \tilde{\sigma}
\left(\nu,Q,a \right) \!
&=& \!
      \exp \Bigg\{ 2\, \int\limits_0^1 \frac{d u}{u} \Bigg[ \,
      \int\limits_{u^2 Q^2}^{u Q^2} \frac{d p_\perp^2}{p_\perp^2}
A\left(\as(p_\perp)\right)
   \nonumber \\   
   & & \qquad  \qquad  \qquad \quad \times \left( {\rm e}^{- u^{1-a} \nu \left(p_\perp/Q\right)^{a} }-1 \right)
\nonumber \\
  + \,\frac{1}{2}   &  & 
   \hspace*{-3.5mm}   B\left(\as(\sqrt{u} Q)\right) \left( {\rm e}^{-u
\left(\nu/2\right)^{2/(2-a)} } -1 \right)
      \Bigg] \Bigg\}. 
\label{thrustcomp}
\ea
The resummation is in terms of anomalous dimensions $A(\alpha_s)$ and $B(\alpha_s)$,
which have finite expansions in the running coupling, $A(\as)=  \sum_{n=1}^\infty A^{(n)}\ (\as/
\pi)^n$,
and similarly for $B(\as)$, with the well-known coefficients  $A^{(1)} = C_F , \, B^{(1)}  =  -3/2 \, C_F 
,\,  A^{(2)} = 1/2 \,C_F [ C_A
( 67/18 - \pi^2/6) - 10/9\, T_F N_f ]$,
at NLL accuracy, independent of $a$, where $C_F = 4/3$, $C_A = 3$, $T_F = 1/2$, and 
$N_f$ denotes the number of flavors. At $a = 0$ we
reproduce the
NLL resummed thrust cross section \cite{CTTW}. 

\section{Universality of Power Corrections}
\label{sec:nonpt}

As it stands, the perturbative cross section, Eq. (\ref{thrustcomp}) is ill-defined for small values of $\tau_a$. This is due to the at best asymptotic nature of the perturbation series, which manifests itself in (\ref{thrustcomp}) as an ambiguity in how to treat the singularities in the running coupling. At $\tau_a \sim \LQCD/Q$ nonperturbative corrections become dominant. Nevertheless, due to the quantum mechanical incoherence of short- and long-distance effects, one can separate the perturbative part from the NP contribution in a well-defined, although prescription-dependent manner (see \cite{shapefn} and references therein). 

Following Refs. \cite{shapefn}, we can deduce the structure of the NP corrections by a direct expansion of the integrand in the exponent at momentum scales below an infrared factorization scale $\kappa$. We rewrite 
 Eq. (\ref{thrustcomp}) as the sum of a perturbative term, labelled with the subscript PT, where all $p_\perp > \kappa$, and a soft term that contains all NP physics:
\ba
\hspace*{-2mm} & \ln & \left[ \frac{1}{\sigma_{\mbox{\tiny tot}}} \tilde{\sigma} \left(\nu,Q,a \right) \right]
 =  2 \left[\int\limits_{\kappa^2}^{Q^2}  \frac{d p_\perp^2}{p_\perp^2} + \int\limits_{0}^{\kappa^2} \frac{d p_\perp^2}{p_\perp^2} \right]  A\left(\as(p_\perp)\right) \nonumber \\
& & \quad \times \int\limits_{p_\perp^2/Q^2}^{p_\perp/Q}
\frac{d u}{u}    \left( e^{- u^{1-a} \nu \left(p_\perp/Q\right)^{a} }-1 \right) 
   + \, B \mbox{-term}
 \nonumber \\
 & \equiv & \ln \left[ \frac{1}{\sigma_{\mbox{\tiny tot}}} \tilde{\sigma}_{\mbox{\tiny PT}} \left(\nu,Q,\kappa,a \right) \right] \nonumber \\
& & + \, \frac{2}{1-a} \sum\limits_{n=1}^\infty \frac{1}{n\, n!} \left(-\frac{\nu}{Q}\right)^n
\int\limits_{0}^{\kappa^2} \frac{d p_\perp^2}{p_\perp^2} p_\perp^n  A\left(\as(p_\perp)\right). 
\label{nonpt}
\ea
We have suppressed terms of order ${\mathcal{O}}(\nu/Q^{2-a} , \nu^{\frac{2}{2-a}}/Q^2 )$, which include the entire $B$-term of Eq. (\ref{thrustcomp}), as indicated. Introducing the shape function as an expansion in powers of $\nu/Q$, from the expansion of the exponent in the second equality of (\ref{nonpt}), with NP coefficients $\lambda_n(\kappa)$,  we arrive at,
\ba 
\tilde{\sigma} \left(\nu,Q,a \right)
& = & \tilde{\sigma}_{\mbox{\tiny PT}} \left(\nu,Q,\kappa,a \right)\, \tilde{f}_{a,\mbox{\tiny NP}} \left(\frac{\nu}{Q},\kappa\right) , \,\,\,\label{ptnp} \\
\ln \tilde{f}_{a,\mbox{\tiny NP}} \left(\frac{\nu}{Q},\kappa\right) & \equiv & \frac{1}{1-a} \sum\limits_{n=1}^\infty \lambda_n(\kappa) \left(-\frac{\nu}{Q}\right)^n.
\ea
We find the simple result that the only dependence on $a$ is through an overall factor $1/(1-a)$ which leads to the scaling rule for the shape function \cite{BS1,BS}: 
\ba
\tilde{f}_{a,\mbox{\tiny NP}} \left(\frac{\nu}{Q},\kappa\right) & = & \left[ \tilde{f}_{0,\mbox{\tiny NP}} \left(\frac{\nu}{Q},\kappa\right) \right]^{\frac{1}{1-a}}. \label{rule}
\ea
Given the shape function for the thrust at $a = 0$ at a specific c.m. energy $Q$, one can predict the shape function and thus from Eq. (\ref{ptnp}) the cross section including all leading power corrections for any other value of $a$. 

Before showing some numerical results, we want to comment briefly on the main assumptions that go into the above derivation. Our starting point is the NLL resummed cross section (\ref{thrustcomp}), which describes independent radiation from the two primary outgoing partons.  Correlations between hemispheres are neglected, although they are present in the resummed formula valid to all logarithmic orders \cite{BKuS}. However, it has been found from numerical studies that such correlations may indeed be unimportant \cite{applshape}. Furthermore, we assume in the separation of perturbative and NP effects (\ref{nonpt}), that long-distance physics has the same properties under boosts as the short-distance radiation. Success of the scaling (\ref{rule}) would thus indicate that NP processes are boost-invariant as well \cite{BS}.

\section{Numerical Predictions}
\label{sec:num}

In \cite{BS} we showed some examples of the shape function in moment space, computed from the perturbative expression (\ref{thrustcomp}), matched to fixed order calculations at NLO with EVENT2 \cite{Catani:1996jh},
and from the output of the event generator PYTHIA \cite{Sjostrand:2001wi} in the absence of data analysis for values of $a$ different from the thrust ($a = 0$). Here we show results in momentum space, after numerical inversion of the Laplace transform.

\begin{figure}
\begin{center}
\epsfig{file=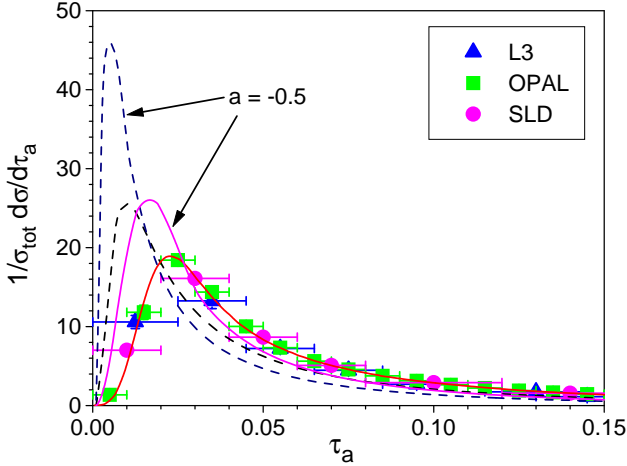,height=8.5cm,angle=270,clip=0}     
\vspace*{-3mm}
\caption{Differential distributions $(1/\sigma_{\mbox{\tiny tot}}) d \sigma/d \tau_a$ for $a = 0$, and $a = -0.5$ at $Q = 91$ GeV. The solid lines are the full cross sections (for $a = 0$ the output of PYTHIA), the dashed lines the perturbative contributions at NLL/NLO. The data are taken from \cite{data}.}
\label{fig:predict}      
\end{center}
\vspace*{-3mm}
\end{figure}
In Fig. \ref{fig:predict} we show our prediction for the differential distribution for $a = -0.5$ in momentum space. This plot is obtained from Eq. (\ref{ptnp}), with the shape function determined via (\ref{rule}) from the one for the thrust, found in \cite{BS}. For further technical details we refer to \cite{BS}.  We also show the differential distributions computed from perturbation theory, and the output of PYTHIA for $a = 0$. As shown, PYTHIA fits the data for the thrust well, thus we take its output to determine the shape function for $a = 0$, instead of fitting a function to the data.

\section{Summary and Outlook}
\label{sec:sum}

We have illustrated how one can test specific properties of long-distance physics with the help of perturbation theory, if supplemented by experimental information. Success or failure of the scaling (\ref{rule}) will provide information about the importance of interjet correlations and about the boost properties of long-range interactions in $e^+e^-$ dijet events. This additional information about nonperturbative physics in turn could, for example, be helpful in precision measurements of $\as$. We hope that experimental tests of our predictions will be carried out in the near future.

\subsection*{Acknowledgements}
\vspace*{-7mm}
\begin{acknowledgement}
We acknowledge many illuminating discussions with Tibor K\'ucs.
C.F.B. thanks the organizers of EPS HEP 2003,
especially Ch. Berger,  V. Chekelian, B. Hirosky, and Z. Trocsanyi,
for the possibility to contribute to a very stimulating meeting, and
the European Union for support at the conference. This work was supported in part by the 
National Science Foundation, grant PHY-0098527.
\end{acknowledgement}

\end{document}